# Demonstration of high-efficiency photonic lantern couplers for PolyOculus

Christina D. Moraitis,[1,*] Juan Carlos Alvarado-Zacarias,[2] Rodrigo Amezcua-Correa,[2] Sarik Jeram,[1] and Stephen S. Eikenberry[1,3,4]

[1]*Department of Astronomy, University of Florida, 211 Bryant Space Science Center, Gainesville, FL 32611, USA*
[2]*College of Optics and Photonics (CREOL), University of Central Florida, 4304 Scorpius St, Orlando, FL 32816, USA*
[3]*Department of Physics, University of Florida, 2001 Museum Rd., Gainesville, FL 32611, USA*
[4]*Institue for High Energy Physics and Astrophysics, University of Florida, 2001 Museum Rd., Gainesville, FL 32611, USA*
*\*c.moraitis@ufl.edu*

**Abstract:** The PolyOculus technology produces large-area-equivalent telescopes by using fiber optics to link modules of multiple semi-autonomous, small, inexpensive, commercial-off-the-shelf telescopes. Crucially, this scalable design has construction costs which are > 10x lower than equivalent traditional large-area telescopes. We have developed a novel photonic lantern approach for the PolyOculus fiber optic linkages which potentially offers substantial advantages over previously considered free-space optical linkages, including much higher coupling efficiencies. We have carried out the first laboratory tests of a photonic lantern prototype developed for PolyOculus, and demonstrate broadband efficiencies of ~91%, confirming the outstanding performance of this technology.

## 1. Introduction

Our team has developed an approach to building large-area-equivalent telescopes by using photonic technology to link multiple semi-autonomous, small, inexpensive, commercial-off-the-shelf (COTS) telescopes. Our scalable design has construction costs which are 10 times lower than equivalent traditional large-area telescopes. This "PolyOculus" array relies on recent advances in low-cost, high-performance COTS equipment - telescopes, CCD cameras, and control computers - combined with a novel photonic-link architecture and a few key technological innovations to produce telescope collecting areas equivalent to standard telescopes with mirror diameters ranging from 0.9-m to 40-m and beyond for certain applications. These include astronomical observations [1,2], space-to-ground optical communications, and atmospheric LIDAR, among others. The basic module consists of a 7-pack of 14-inch COTS telescopes which are semi-autonomous in their pointing, tracking, and focus control, equipped with an acquisition/guiding system and a fiber optic feed. The output of this 7-pack can be sent directly to the detector system, or can be linked to other 7-packs in a hierarchical, scalable architecture to combine the light before sending it to the final detector.

The initial PolyOculus concept used a free-space optical relay to link the outputs of the 7 individual telescopes into a single fiber [1]. This design used a collimator/camera optical lens relay to re-image the output of the 7 individual fibers, arranged in a heptagonal bundle, onto the input of a single fiber, with the collimator/camera focal lengths selected to provide the desired magnification needed to match the diameter and numerical apertures of the heptapack bundle to the single output fiber. In addition to the optical throughput challenges of this relay (with ~80% efficiency), the fill factor of the fiber packing at the relay input has an important impact on PolyOculus "downstream" performance [1]. Any packing efficiency <100% requires that the etendue (the area times solid angle, or "*AΩ* product") of the next optic in the system increase by 1/(fill factor) to avoid loss of light. For spectrographs, this will increase the required pupil image size (and thus require larger, more expensive gratings and collimator/camera optics) [3]. For next-level PolyOculus combiners, this can eventually lead to fibers at the limit of current technology in terms of etendue. For circular 7-packs inside a circular aperture, the maximum packing theoretical efficiency is ~78% (in the extreme and impractical limit of zero cladding thickness). Another approach we previously considered uses "pupil plane packing", which avoids the drive to zero cladding thickness. In this approach, the 7 input fibers are each coupled to a lenslet array creating a pupil plane with the pupil images closely packed together. The lenslet array is designed to maximize the fill factor of the pupil plane - the individual pupil image diameters match the lenslet pitch in the array, which in turn matches (or slightly exceeds) the clad diameter in the input fiber plane. While the fiber clad diameter thus influences the lens array, this approach retains flexibility via the focal length of the lenslets, which is straightforward to adjust to match the desired design. Furthermore, the packing efficiency of lenslets can still be relatively high – several reputable vendors report regularly achieving ~75% (close to the theoretical maximum of 78% for circular packing above).

An alternative, and potentially superior, approach is to couple the 7 optical fiber feeds of the telescope units to a single output fiber using a photonic lantern [4-7]. Photonic lanterns adiabatically couple input fibers into a single output fiber (or vice versa), with the potential inputs ranging from single-mode to few-mode, or even multi-mode fibers, and the output being a multi-mode fiber. Previous photonic lanterns have been pushed to coupling efficiencies with >97% broadband transmission for multimode inputs and outputs (like PolyOculus will use). The photonic lantern also preserves etendue (expressed as mode-matching - see below) with efficiencies approaching 100%. These features make the photonic lantern an ideal match to the needs of PolyOculus, where efficiency and etendue maintenance are both at a premium, and represent a substantial improvement over our previous free-space optical coupler designs above.

In this paper, we present the first throughput measurements of a photonic lantern designed to meet the requirements of PolyOculus. In Section 2 below, we describe the photonic lantern design and the fabrication procedures for the test lantern we studied here. In Section 3, we describe the efficiency testing setup and procedure, and the resulting efficiencies we determined for white light and various bandpass filters in the optical (400-1100 nm) waveband. Finally, in Section 4 we present our conclusions and directions for future work.

## 2. Photonic lantern design and fabrication

### 2.1 Photonic lantern design parameters

We designed the photonic lantern used in this experiment to couple light from 7 individual fibers into a multimode delivery fiber at the output of the photonic lantern. At the wavelength range of interest (400-900 nm), the initial input fibers support ~4 to ~20 modes, and are effectively Few Mode Fibers (FMFs). An adiabatic taper maximizes the efficiency of the photonic lantern, efficiently mapping the modes of the bundle of fibers at the input into the multimode output of the photonic lantern. Fig. 1 (a, b) shows the refractive index profile of the initial structure of the photonic lantern for numerical analysis, resembling those of the actual fabricated photonic lantern. 7 individual fibers are stacked together into a fluorine-doped tube with a refraction index difference of -18x10$^{-3}$ with respect to silica.

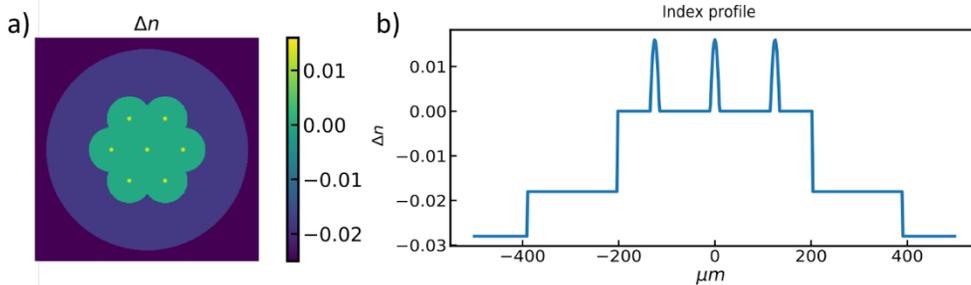

Fig. 1. Design parameters for the fabricated photonic lantern. (a) 2D refraction index profile showing the core positioning of the designed photonic lantern inside the fluorine doped tube. (b) 1D index cross-section profile of the simulated photonic lantern.

Fig. 2 (a) shows the end-facet view of the fabricated photonic lantern and Fig. 2 (b) shows the mode evolution as the photonic lantern is tapered. At the beginning of the plot, when the photonic lantern core is around 60 μm (0.16 taper ratio - the ratio of the output diameter to the input diameter), the individual cores in the photonic lantern are considered isolated as they are placed far apart and their normalized propagation constants are expected to be the same as indicated by the overlapping of the 7 propagation constants at the beginning of the taper (red lines). As the whole structure is scaled down, the initial 7 supermodes [8] in the structure evolve into the modes of the multimode waveguide at the end of the photonic lantern taper, as indicated by the splitting of the propagation constants at around 30 μm (0.08 taper ratio). The propagation constants do not cross along the taper, indicating that the taper is adiabatic (meaning that the transmission losses and intermode couplings are theoretically zero [9]) and each propagation constant evolves into a specific mode at the multimode end. Note that for a perfectly adiabatic taper, the transmission has 100% efficiency and the etendue of the output light in free space is perfectly conserved. The blue lines represent the cladding modes supported by the waveguide and we clearly see there is no crossing with the core modes, indicating there is no power transfer and the core modes stay within the photonic lantern core.

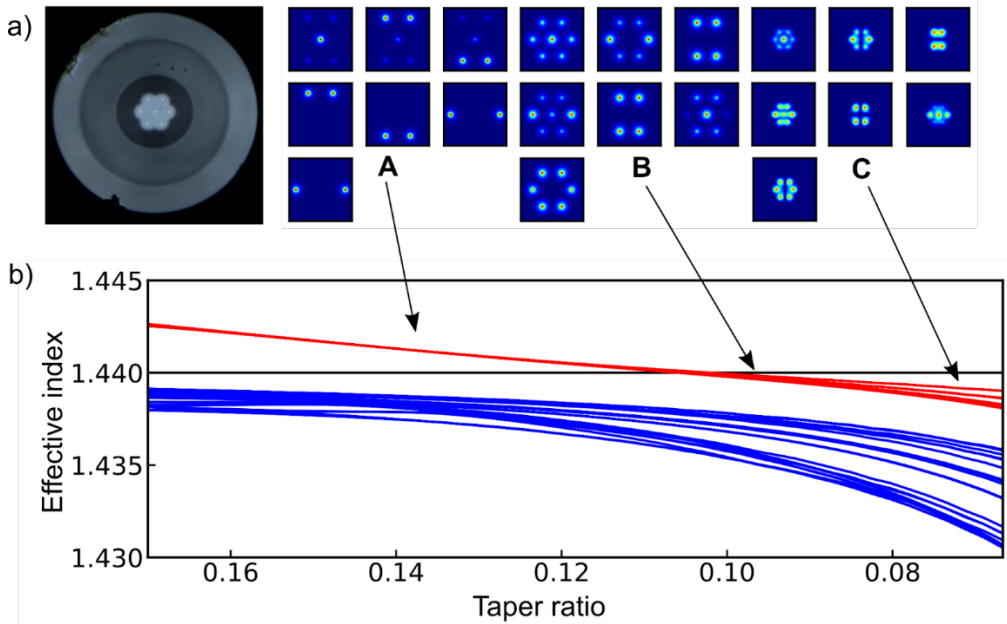

Fig. 2. (a) Optical microscope image of the fabricated photonic lantern. (b) Effective index of refraction evolution along the taper, where the red lines correspond to the effective index of the 7 initial fibers and the blue lines correspond to the cladding modes in the photonic lantern. In A the light starts to couple between cores and as the photonic lantern is tapered they become more coupled and as seen in B, and in C, shows the mode profiles supported by the multimode end of the photonic lantern. The propagation constants stay together along the taper and close to the taper end they start to split up and evolve into the modes supported at the end of the taper.

We numerically analyze the behavior of the photonic lantern taper using the beam propagation method (BPM)[10] using a suite of Python code developed at CREOL for this purpose. In Fig. 3a we can see the initial supermodes supported by the photonic lantern structure when the core is around 50 $\mu$m, these initial modes are then propagated using BPM over a distance of 25 mm until the final core of the photonic lantern is 20 $\mu$m, the output modes after BPM are presented in Fig 3b, where we observe some of the modes resemble those of the multimode end of the photonic lantern (see Fig 3c), while others present some mode crosstalk within the same mode group, especially for higher order modes. While mode crosstalk can present a challenge for photonic lantern applications in data transmission, it has no negative impact on PolyOculus power transmission, since in data transmission mode crosstalk from all supported modes within the multimode fiber respect one specific mode impairs the transmission through such fibers [11-13].

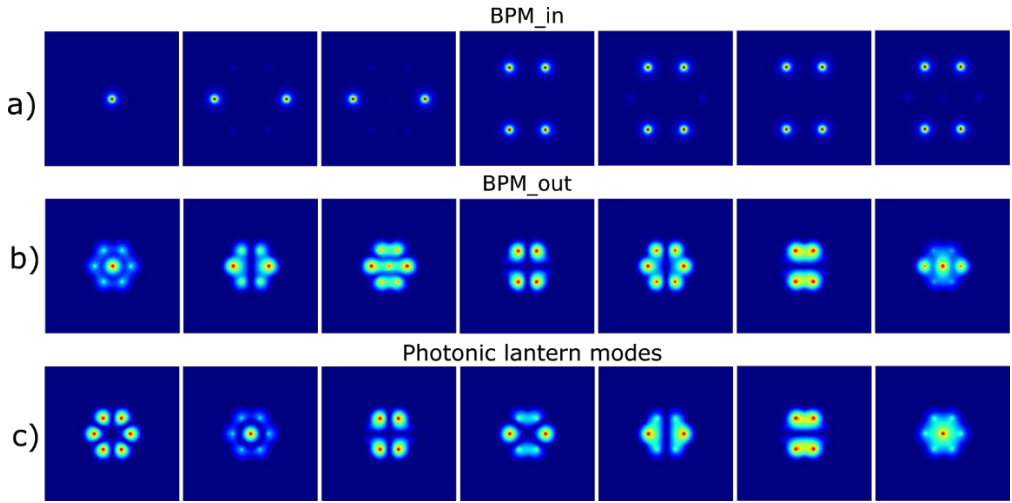

Fig. 3. Beam propagation simulations through photonic lantern, a) initial supermodes launched into the photonic lantern. b) Modes at the end of the photonic lantern after BPM. c) Modes supported at the end of the photonic lantern structure without BPM simulation.

### 2.2 Photonic lantern fabrication

The photonic lantern fabrication closely follows the design developed in the previous section. The initial input fibers are step index fibers with 8.2/125 $\mu$m core and cladding diameters respectively and 0.14 numerical aperture (NA). These fibers are subsequently spliced to a short piece (5cm) of graded index fiber having a core/cladding size of 19/125 $\mu$m respectively with an index difference of $16 \times 10^{-3}$. The graded index fibers are then inserted into a low refractive index capillary that will be adiabatically tapered to the multimode end of the photonic lantern. The use of graded index fibers eases the adiabaticity requirements of the fabricated photonic lantern [14, 15], providing low loss over the photonic lantern transition. Once the fibers are inserted into the low refractive index capillary, we adiabatically taper the photonic lantern in two steps using a $CO_2$ laser-based tapering station. In the first step, we taper the initial capillary by a factor of 2 over a 1.5 cm taper length and in the last step, we taper the photonic lantern by a factor of 7.2. The final photonic lantern device has a core diameter of 25 $\mu$m with a cladding of 115 $\mu$m, and a calculated NA of ~0.22.

As a preliminary measurement of the coupling efficiency from the input fibers to the multimode end of the photonic lantern, we coupled light from a laser source centered at 1064 nm into each of the input fibers, one at a time, and measured the output power at the photonic lantern end-facet using an optical power meter (Thorlabs S144A). We summarize the coupling efficiencies for each of the 7 photonic lantern ports in Table 1. While the high (~98%) efficiency achieved in this test is promising, the light source, injection, and modal properties here differ enough from the PolyOculus use case that we carried out an independent test for PolyOculus-like efficiency at the University of Florida, which we describe in the next section.

Table 1. Measured coupling efficiency through the photonic lantern with a 1064 nm laser.

| Core number | 1 | 2 | 3 | 4 | 5 | 6 | 7 | Average |
|---|---|---|---|---|---|---|---|---|
| Coupling Efficiency (%) | 97.2 | 97.2 | 96.8 | 97.8 | 97.6 | 98.5 | 98.7 | 97.7 |

## 3. Efficiency testing and results

### 3.1 Experimental setup

In order to measure the optical throughput efficiency of the photonic lantern for PolyOculus, we created a laboratory test setup which mimics the PolyOculus use case. In PolyOculus, each unit telescope creates an image of a target star at its focal plane, with a typical diameter of ~19 microns (full-width at half-maximum) at an f/10 beam speed (NA = 0.05). An optical relay then re-images this star image onto the photonic lantern input fiber at NA=0.12 with a corresponding ~8-micron image diameter.

To simulate this in the laboratory environment, the testing apparatus consists of an illumination chain (Fig. 4) that injects light into the input fiber being tested, the output flux of which is measured using a ThorLabs S130C power meter (Fig 5a). We made throughput efficiency measurements of the prototype photonic lantern for PolyOculus using a white light source with an approximate bandpass from 400-900nm, and an effective average wavelength of ~650nm. Light from the lamp is injected into a multimode fiber, the output of which is placed at the input focal plane of the illumination chain (Fig. 4, "Light Source"). The light passes to a collimating lens (Collimator 1) which creates a pupil image with an aperture stop to control the beam speed of the light which is eventually injected into the photonic lantern input fibers. The beam then reflects off of a fold mirror (due to spatial constraints on our test bench) and passes to a re-imaging lens (Camera 1). We place a 300-micron diameter pinhole at the focal plane created by Camera 1, to constrain the beam size. After this, we employ another collimator/camera optical relay (Collimator 2 and Camera 2 in Fig. 4) to place another focused image of the pinhole at a focal plane accessible to the input fibers. We seat the fibers into brass fiber chucks (eight in total - 7 for photonic lantern input fibers and 1 for a reference fiber) that are secured onto an 8-slot fiber chuck holder by a cover plate (Fig 5). This chuck holder is installed onto a custom adjustable tip-tilt stage, actuated by spring-loaded screws. This tip-tilt stage is mounted onto a custom-designed 3D-printed stage that travels in the focus direction (Z), driven by a Schneider Electric motor. This focus stage is installed onto a Physik Instrumente (PI) M-403 linear stage that travels in the vertical direction (Y) and this vertical stage itself is installed on to another PI M-403 linear stage that travels in the horizontal direction (X), allowing for translation in X, Y and Z. We refer to this assembly of horizontal, vertical, and focus direction stages as the translation stage and show it in Figure 5.

Finally, we place the Thorlabs S130C power meter at the output end of the photonic lantern to measure the output flux using ThorLabs Power Meter software (Fig. 5a). We call this the "front-illuminated" configuration of our test system. We now describe the procedure used to carry out the photonic lantern efficiency measurements.

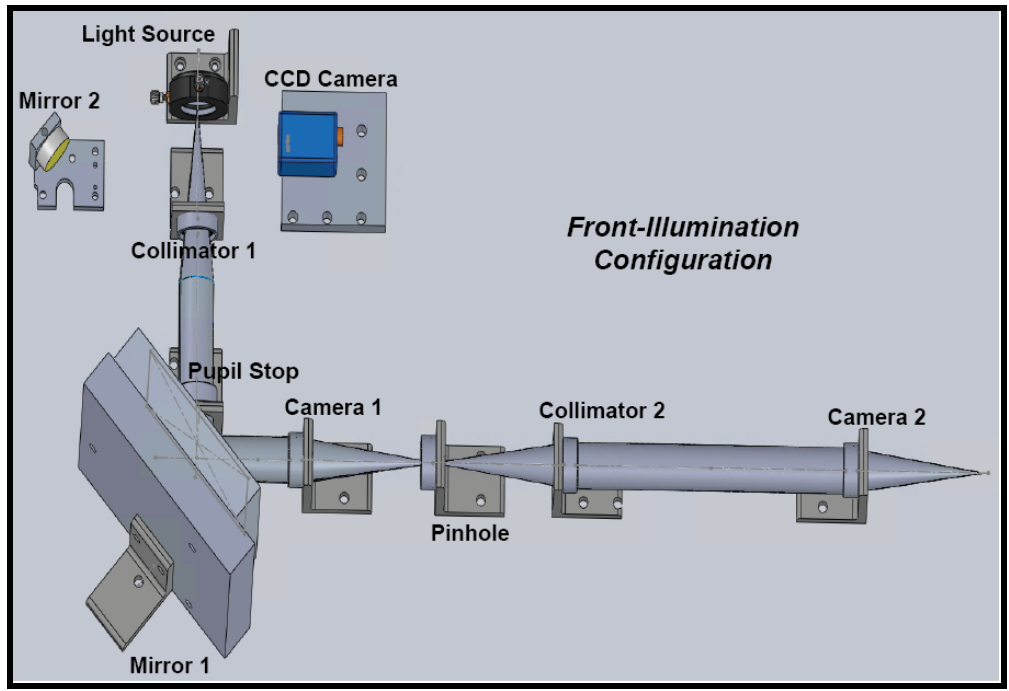

Fig. 4. Overhead view of front-illuminated configuration of the system. The model was created in SolidWorks and the beam path in Zemax. Light passes through the optical configuration from an optically aligned light source and is focused down to a point at the fiber input beyond Camera 2. This light enters into the fiber input end and passes through the photonic lantern to the photonic lantern output where a Thorlabs power meter reads the flux in nW seen in the extended view in Figure 5a.

## 3.2 Experimental procedure

We carry out the efficiency measurements using this setup in a two-step process, front and back-illumination. We first back-illuminate the output end of a fiber in order to align the input fiber correctly in X, Y, and Z, as well as tip-tilt. The back-illuminated configuration injects light into the nominal "output" end of the photonic lantern or reference fiber (Fig. 5b), which then propagates backward through the system (Fig. 6). Prior to the front-illuminated injection source, we insert a fold mirror (mirror 2 in Fig. 6) into the beam, directing the back-illuminated light into a ImageSource DMK 23U445 CCD camera.

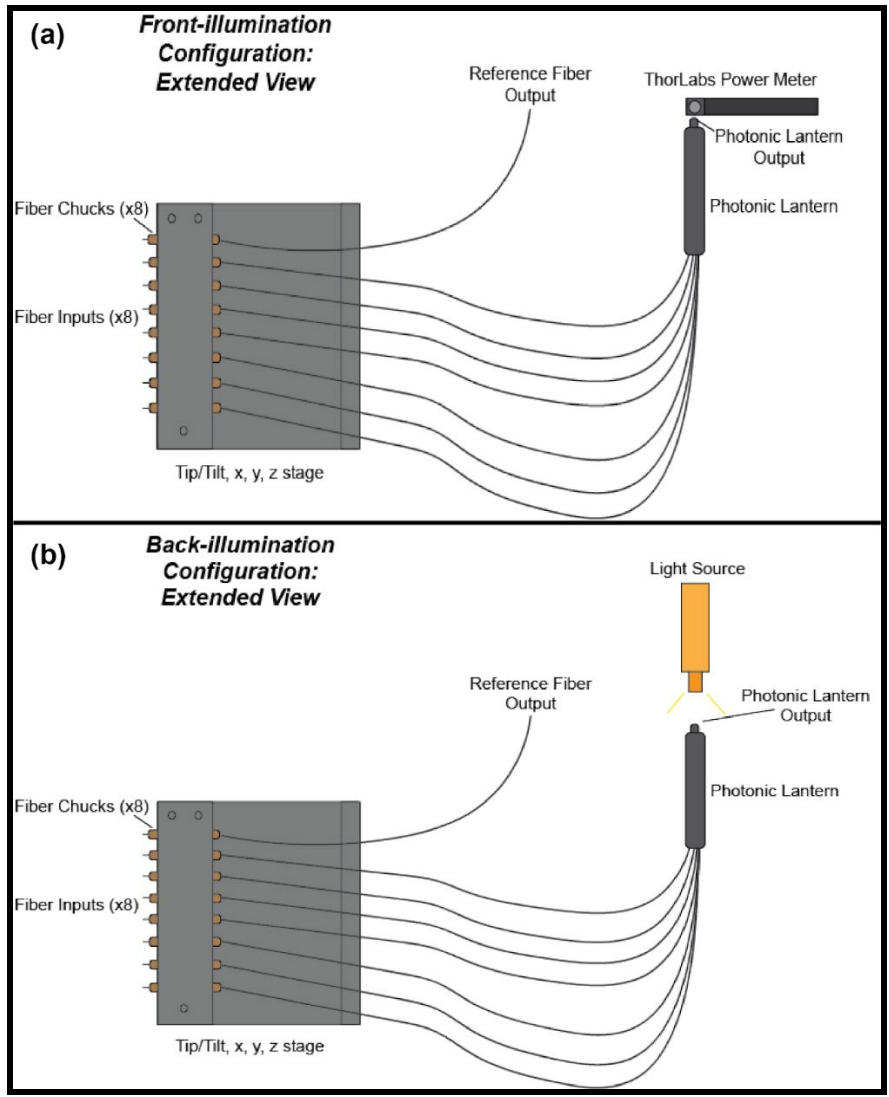

Fig. 5. Overhead view of extended back/front illumination configurations of the system. This configuration represents what lies beyond Camera 2 from Fig. 4 and Fig 6. For front-illumination (a), light from the illumination chain (Fig. 4) enters into the fiber input end and passes through the photonic lantern to the photonic lantern output where a Thorlabs power meter reads the flux in nW. For back-illumination (b), light enters the output end of the PL and exits through the input ends and travels backward through the illumination chain (Fig. 6) to a CCD camera for alignment. We number the fibers 1-7, starting at the bottom, and ending with the reference fiber at top. The reference fiber is the same type of fiber used in all 7 fiber inputs of the photonic lantern configuration.

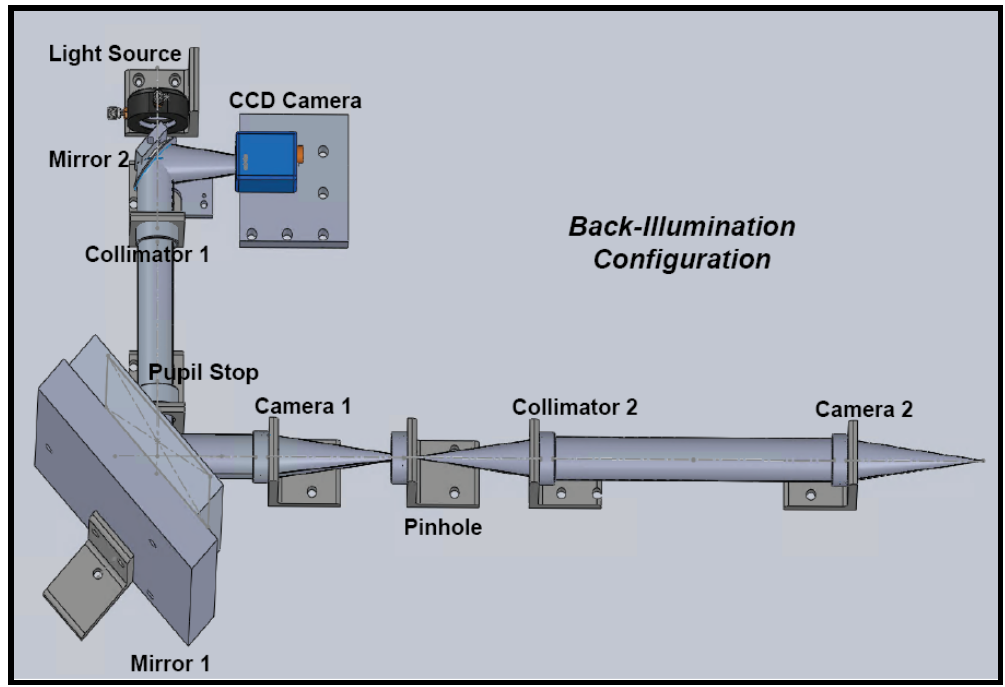

Fig. 6. Overhead view of back-illuminated configuration of the system. The model was created in SolidWorks and the beam path in Zemax. Light is injected into the output end of the photonic lantern (extended view seen in Fig. 5b) and comes out the input ends of the fibers (beyond camera 2). The light beam passes backwards through the optical configuration, this time meeting a fold mirror after collimator 1, to a ImageSource DMK23U445 CCD camera where we align the beam through the 300 micron pinhole.

We used the CCD to ensure the beam passes through the pinhole by adjusting the translation stage such that the beam is centered at the pinhole center. Once the beam was centered, we optimized the beam's focus by adjusting the translation stage in the focus direction. Before adjusting the tip and tilt stage, we imaged the pupil plane (located at the pupil stop in Fig. 6) by installing a lens on the CCD. The tip-tilt stage was adjusted accordingly to ensure uniform pupil plane illumination. Throughout our testing, tip and tilt adjustments were not often required, as the fiber tips in the apparatus were sufficiently co-aligned.

Once we obtained uniform pupil plane illumination and co-alignment of the pinhole image with the fiber ends, we shifted to the front-illumination configuration to obtain flux measurements at the output end of the fibers. Working with such small fibers presented challenges in precise repeatability, from movement due to thermal expansion/contraction of the aluminum setup, fluctuations of the light source over long periods of time, and possibly other effects. To counter this, we alternated between measurements of the reference fiber and the photonic lantern input fibers, so that each photonic lantern fiber had a reference measurement taken close to it in time. The reference fiber is identical to all input fibers of the photonic lantern, and we use it to determine the expected power transmitted without a PL. For each fiber (reference or photonic lantern), we began with white light power measurements. We then apply small offsets in X, Y, and Z to maximize the detected power. Once we reach a stable maximum power position, we recorded 100 data points per fiber flux measurement using the ThorLabs Power Meter software (1 data point per second for 100 seconds). Typical flux values ranged from 105-118 nW in white light over the course of measurement. We measured any excess light from the apparatus (with the lamp switched off) to be ~0.01 nW, and thus negligible. Once the flux measurements were recorded, we returned back to the back-illumination process with a different fiber and repeated until completion.

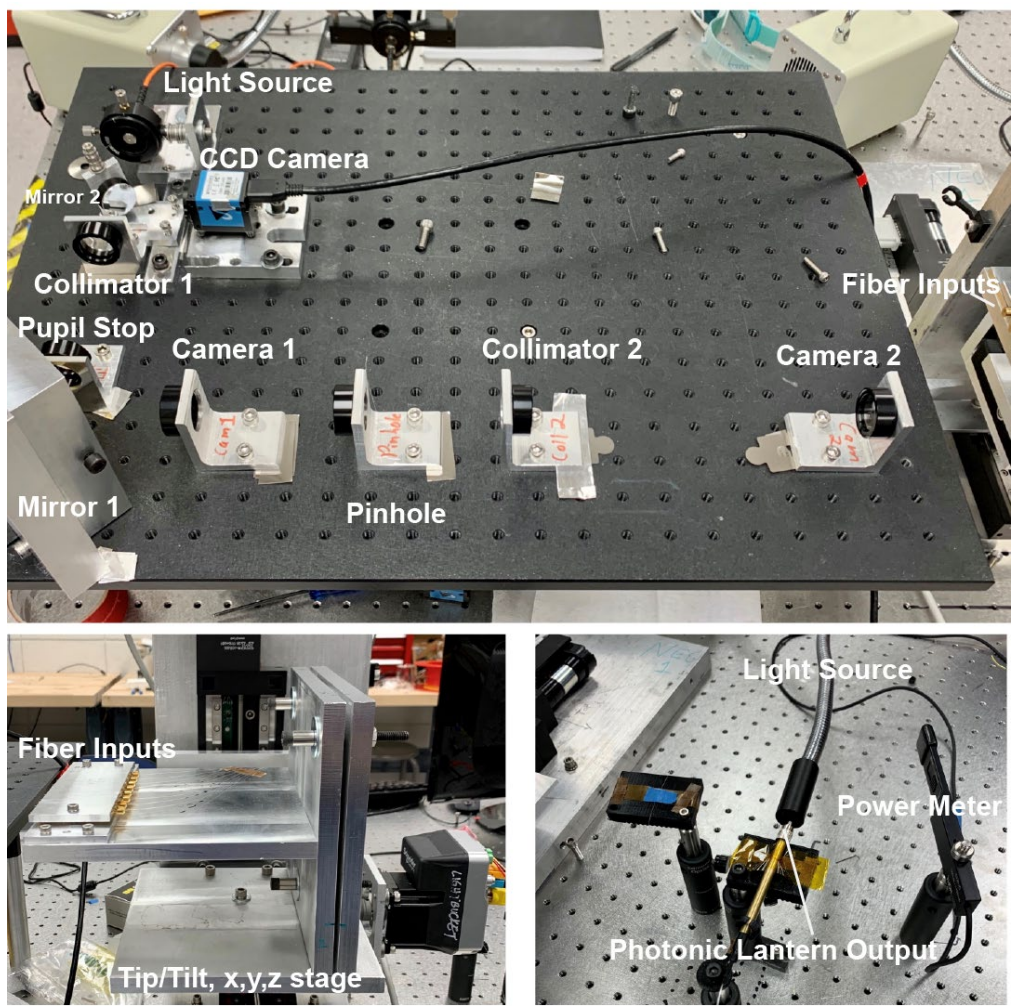

Fig. 7. (Top) An overhead of the optical bench, similar to Figure 6, shows how the light beam travels through the optical path. Similar to the model seen in Fig 6, this specific image shows back-illumination with the pinhole mounted properly and the fold mirror screwed in place before the ImageSource CCD camera. (Bottom left) The fibers are held in separate, individual chucks and mounted onto the X, Y, Z / tip-tilt stage. The stage moves using PIMikroMove electronic controls (X and Y) and Schneider Electric motor. (Bottom right) The photonic lantern output and reference fiber output lie to the right and below the X, Y, Z / tip-tilt stage. During back-illumination, the lamp is switched on and placed in front of the output end (of either the photonic lantern or reference fiber depending on the fiber being measured). During front illumination, the Thorlabs power meter is placed in front of the output end of the fiber of interest.

### 3.3 Efficiency results

For the current photonic lantern prototype for PolyOculus, we obtained results from 6 of the 7 photonic lantern fibers, interleaved with reference fiber measurements. (Fiber 3 was damaged during packaging prior to shipment at the University of Florida. While we are able to repair broken photonic lantern fibers effectively using splicing, time and shipping constraints prevented us from doing so for this paper, and we exclude Fiber 3 from the analysis.) We found that the variation of the 100x 1-second measurements was notably small, with an RMS well below 0.01-nW. Over the course of previous tests with this setup, we measured a number of fibers using this procedure, and we found the long-term repeatability of flux measurements to have an RMS of ~1.5%, which we adopt as our overall uncertainty. Given the extremely small contributions of background light and variation in the 100-second power readings, we conclude that the dominant source of error is likely to be variations in the optical alignment, which in turn produce variations in the launch of light from the illumination system into the fiber. We note also that we see the same variations in both PL and reference fibers, and these persist for measurements made on different days and under different experimental setups. Thus, we believe that this 1.5% variation takes into account all of the non-repeatability of the system. However, the 1.5% level of uncertainty is sufficient for carrying out the efficiency measurements here.

We present our primary efficiency results in Table 2, over two independent iterations of the efficiency measurements carried out several days apart. The photonic lantern consistently shows white light efficiency of ~90%, with average and median values of ~91% across fibers and iterations.

Table 2. Efficiency values of each photonic lantern fiber taken on two separate days.

| Photonic lantern fiber number | Iteration 1 (% efficiency) | Iteration 2 (% efficiency) | Average (% efficiency) |
|---|---|---|---|
| 1 | 89.9 ± 1.5 | 91.1 ± 1.5 | 90.5 ± 1.1 |
| 2 | 87.9 ± 1.5 | 90.4 ± 1.5 | 89.2 ± 1.1 |
| 4 | 90.4 ± 1.5 | 87.9 ± 1.5 | 89.2 ± 1.1 |
| 5 | 92.1 ± 1.5 | 92.8 ± 1.5 | 92.5 ± 1.1 |
| 6 | 93.0 ± 1.5 | 92.0 ± 1.5 | 92.5 ± 1.1 |
| 7 | 91.3 ± 1.5 | 91.2 ± 1.5 | 91.3 ± 1.1 |
| Average | 90.8 ± 0.6 | 90.9 ± 0.6 | |
| Median | 90.9 ± 0.6 | 91.1 ± 0.6 | |

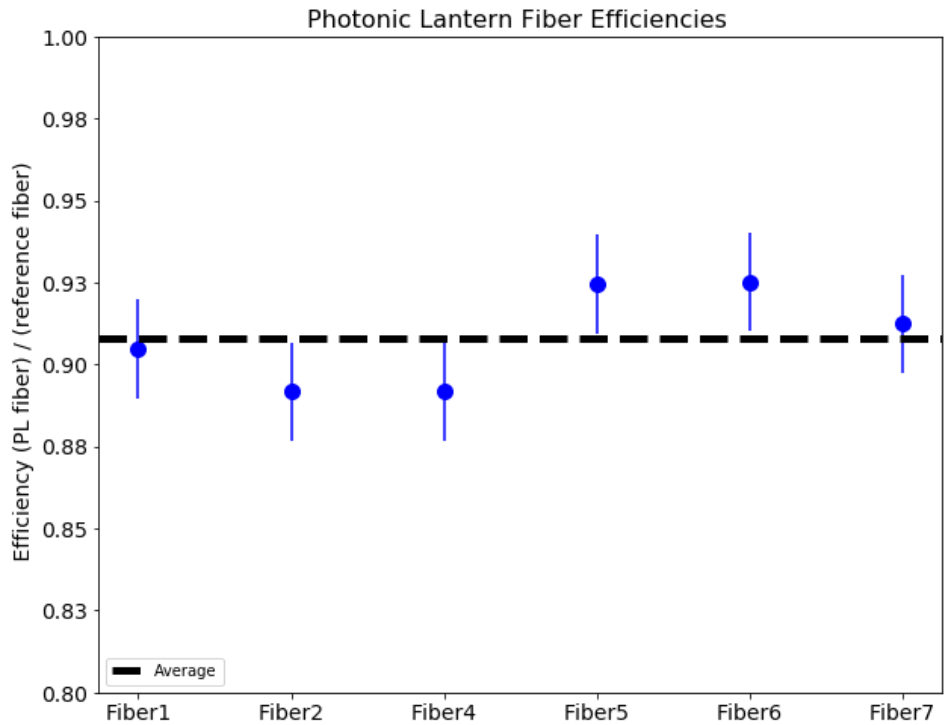

Fig. 8. White light efficiency measurements for the photonic lantern fibers. The horizontal dotted line indicates the average value of ~91% efficiency, for reference. We calculated the efficiency by taking the ratio of the lantern fibers' flux to the flux of reference fibers. Note that Fiber 3 is excluded from the analysis. The values given in the figure reflect the average efficiency value between two separate iterations of measurements (column 4 in Table 2) with an uncertainty of ± 1.5.

## 4. Conclusions

We have shown that a prototype Photonic Lantern for PolyOculus, using the nominal fiber parameters needed for coupling light from the telescope units, can combine light with a throughput efficiency of ~91% in white light. Furthermore, due to the inherently adiabatic nature of the photonic lantern design, it has near-perfect preservation of etendue (versus geometric approaches <80% - see section 1). Any practical degradation in etendue is reflected in the throughput efficiency, as the photonic lantern effectively filters out those modes of propagation. This confirms the utility of photonic lantern couplers for PolyOculus - a significant technological step forward for that technology.

In future work, we plan to characterize the efficiency of the PolyOculus photonic lantern coupler as a function of wavelength. We will also measure the beam quality and etendue, particularly as it relates to injection of the photonic lantern outputs into a standard astronomical spectrograph.

**Acknowledgments.** The authors would like to thank Dr. S. Leon-Saval for making the key introductions that led to this work. SJ acknowledges the support of a University of Florida Graduate Student Fellowship. SSE and CDM were supported in part by Elizabeth Wood Dunlevie Honors Term Professorship at the University of Florida.

**Funding**

**Data availability.** Data underlying the results presented in this paper are not publicly available at this time but may be obtained from the authors upon reasonable request.